\documentstyle[12pt,aaspp4,astrobib]{article}

\def\Cfour{\protect\ion{C}{4}}
\def\Cfourwave{\protect\ion{C}{4} $\lambda \lambda 1548, 1551$}

\def\ERGS{ergs cm$^{-2}$ s$^{-1}$}
\def\INTENS{ergs cm$^{-2}$ s$^{-1}$ sr$^{-1}$}
\def\HA{H$\alpha$}

\def\LA{Ly$\alpha$}
\def\LB{Ly$\beta$}

\def\Osix{\protect\ion{O}{6}}
\def\Osixwave{\protect\ion{O}{6} $\lambda \lambda 1032, 1038$}

\def\SLITsix{10\arcsec $\times$56\arcsec}
\def\SLITseven{20\arcsec\ diameter}

\def\cf{cf.}
\def\ebv{$E($\bv)}
\def\eg{e.g.}
\def\etal{et al.}
\def\euve{{\it EUVE}}
\newcommand{\fig}[1]{Fig.~\ref{#1}}

\def\hut{HUT}
\def\ie{i.e.}

\def\rosat{{\it ROSAT}}

\newcommand{\sig}[1]{#1~$\sigma$}

\begin{document}

\title{Limits on Far-UV Emission from Warm Gas in Clusters of Galaxies
 with the Hopkins Ultraviolet Telescope}


\author{W. Van Dyke Dixon and Mark Hurwitz}
\affil{Space Sciences Laboratory, University of California\\
Berkeley, California 94720\\
vand@ssl.berkeley.edu, markh@ssl.berkeley.edu}

\and

\author{Henry C. Ferguson}
\affil{Space Telescope Science Institute, Baltimore, Maryland 21218\\
ferguson@stsci.edu}

\begin{center}{To appear in {\it The Astrophysical Journal (Letters)}}\end{center}

\begin{abstract}

We have searched the far-UV spectra of five clusters of galaxies
observed with the Hopkins Ultraviolet Telescope (HUT) for emission in
the resonance lines of \Osixwave\ and \Cfourwave. We do not detect
significant emission from either species in any of the spectra.
Lieu \etal\ [ApJ, 458, L5 (1996)] have recently proposed 
a warm [$(5-10) \times 10^5$~K] component to the intracluster medium (ICM)
to explain excess 0.065--0.245 keV flux
present in \euve\ and \rosat\ observations of the Virgo cluster.
If the surface brightness of this warm component follows that of the hot,
x-ray-emitting gas (\ie, is centrally condensed),
then our upper limit to the \Osix\ surface
brightness in M87 is inconsistent with the presence of substantial $5
\times 10^5$~K gas in the center of the Virgo cluster. This inconsistency
may be alleviated if the central gas temperature is $\gtrsim 7.5 \times
10^5$~K. HUT limits on the \Osix\ surface brightness of the four other
clusters can provide important constraints on models of their ICM.

\end{abstract}


\clearpage

\section{Introduction}
 
X-ray observations of clusters of galaxies have long revealed the
presence of hot ($2 \times 10^7$~K), low-density gas permeating many
clusters \cite{FJ82}. Recently, \citeN{LMBLHS96} have
reported the detection of extended, extreme-ultraviolet (EUV) emission
in the Virgo cluster. The emission, observed in the 0.065--0.245 keV
energy band with the {\it Extreme Ultraviolet Explorer} (\euve),
extends to a radius of some 20\arcmin\ about M87, the central galaxy of
the Virgo cluster. The excess is also present in archival data obtained
with the \rosat\ PSPC \cite{Boh95}, which overlaps the \euve\ wavelength
band at low energies. The authors interpret this
emission as evidence for a second, warm [$(5-10) \times 10^5$~K]
component to the intracluster medium (ICM).

The warm gas, with a total mass of approximately $8.9 \times 10^{10}
M_{\sun}$, would cool at a rate of at least 340 $M_{\sun}$ $yr^{-1}$,
more than 30 times the cooling-flow rate deduced from X-ray
observations of the hot ICM \cite{SCFN84}. This discrepancy has
led \citeN{Fabian96} to suggest that the apparent EUV excess is an
artifact of the interstellar absorption model used in the data
analysis. EUV flux is strongly attenuated by gas in our Galaxy's
interstellar medium (ISM). If the Galactic absorption is slightly less
than currently predicted, then an EUV excess is not required to
reproduce the observed spectrum.

In a note added in proof, \citeANP{LMBLHS96} report that {\it ROSAT}
data of Coma and the Abell clusters 2199, 1795, and 1367 are better
modeled by a two-temperature gas in which the cooler component has $T
\sim 10^6$~K than by a single-temperature hot gas, indicating
that warm gas may not be limited to the Virgo cluster. If a reservoir
of warm gas were present in the cores of clusters of galaxies, it would
emit strongly in the far-UV (FUV) resonance lines of \Osixwave\ and
\Cfourwave\ as it cools through temperatures of a few times $10^5$~K
\cite{EC86,VDS94}. Limits on the
intensity of FUV lines from clusters of galaxies can thus provide
important constraints on the mass of the warm component of the ICM. To
this end, we have used the Hopkins Ultraviolet Telescope (HUT) to
search for FUV line emission from clusters of galaxies.

\section{Observations and Data Reduction}

Our observations were carried out with HUT on the Astro-2 mission of
the space shuttle {\it Endeavour} in 1995 March. HUT consists of a
0.9~m, f/2 mirror that feeds a prime-focus spectrograph with a
microchannel-plate intensifier and photodiode array detector.
First-order sensitivity extends from 820 to 1840 \AA\ at 0.51
\AA\ pixel$^{-1}$. The resolution is about 4~\AA\ through the
spectrograph's \SLITsix\ aperture and 7~\AA\ through its
\SLITseven\ aperture. The effective area is nearly 24
cm$^2$ at 1032~\AA. The spectrograph and telescope are described by
\citeN{HUT_INSTR}, while \citeN{HUT_INSTR2} discuss modifications,
performance, and calibration for the Astro-2 mission.

Of the five clusters of galaxies, Virgo, Coma, and the Abell clusters
2199, 1795, and 1367, for which \citeANP{LMBLHS96} find evidence of
$(5-10) \times 10^5$~K gas, all but Abell 2199 were observed with HUT
on Astro-2. The Hercules cluster (Abell 2151) was also observed and is
included in our sample. The observations were obtained during orbital
night, minimizing the strength of the various airglow features and the
scattered light from \LA, which is quite strong during the day. Two of
the targets, Coma and Hercules, were observed twice; their individual
spectra were added together for our analysis. For neither Coma nor
Hercules was the telescope pointed at a particular galaxy. The two
Coma pointings were offset from one another by some 2\farcm4, roughly
the resolution of the model that \citeANP{LMBLHS96} fit to the EUV
data. The M87 pointing was centered on the galaxy's center of light.
Table \ref{targets} lists the observations in our sample.

In our analysis, models are fit to raw-counts spectra using the
non-linear curve-fitting program SPECFIT \cite{Kriss94}, which runs
in the IRAF\footnote{ The Image Reduction and Analysis Facility (IRAF)
is distributed by the National Optical Astronomy Observatories, which
is operated by the Association of Universities for Research in
Astronomy, Inc. (AURA) under cooperative agreement with the National
Science Foundation.} environment and performs a $\chi^2$ minimization.
Error bars are assigned to the data assuming Poisson statistics.
Emission line intensities are converted to energy units using the HUT
absolute calibration, which is based on white dwarf model atmospheres
and is believed accurate to about 5\% \cite{HUT_INSTR2}. Throughout
this paper, line intensities are expressed in units of \INTENS\ and
refer to the sum of the doublet line intensities.

\begin{figure}
\plotone{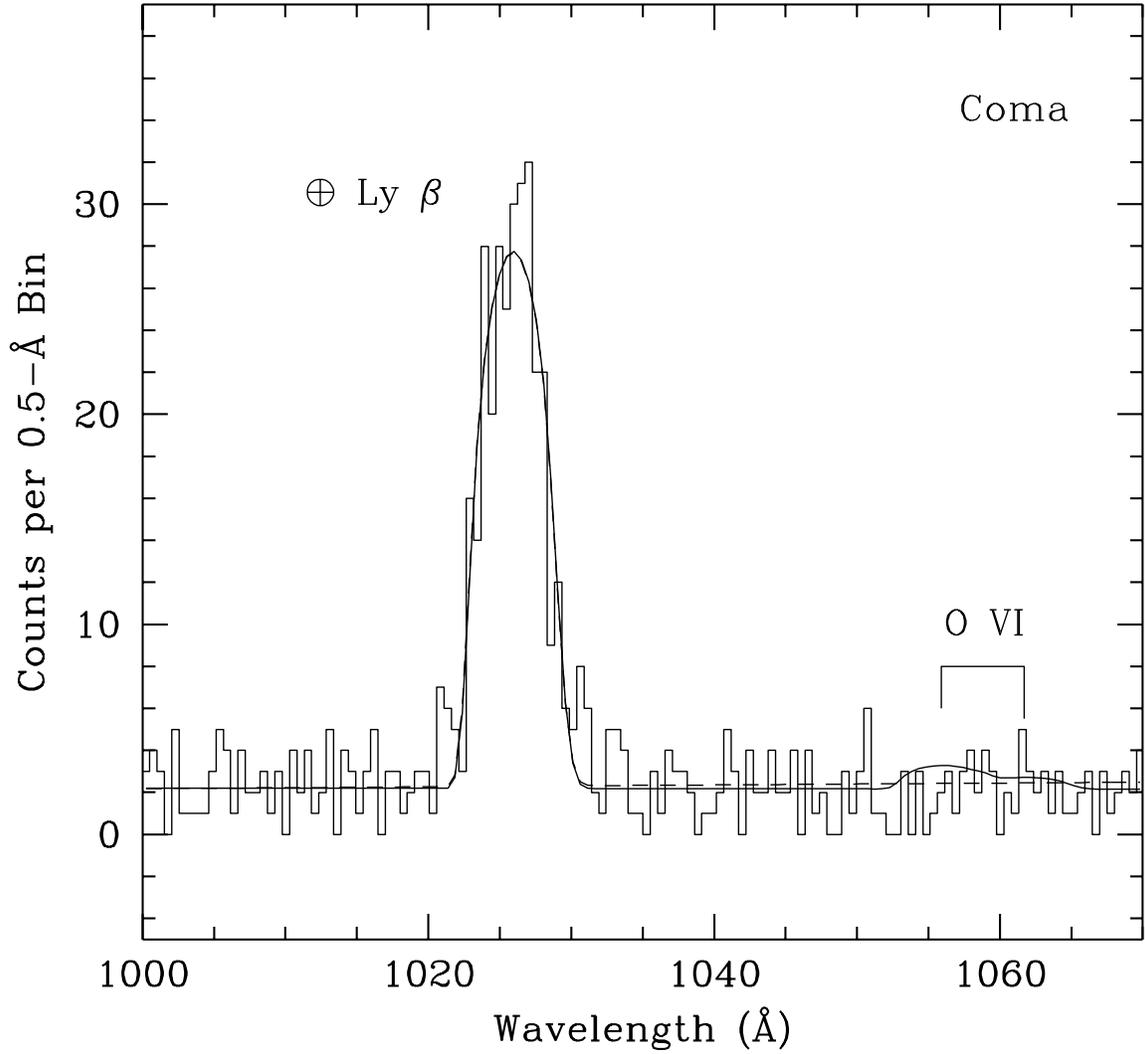}
\caption{HUT spectrum of the Coma cluster, showing the
region about redshifted \Osixwave. The data are shown as a histogram
and are overplotted by models including no \Osix\ emission (dotted
line) and \Osix\ at our 2 $\sigma$ upper limit (solid line).
\label{coma}}
\end{figure}

To set limits on the observed \Osix\ emission, we fit a linear
continuum and three emission features, the \LB\ $\lambda 1026$ airglow
line and the redshifted \Osixwave\ doublet, to a segment of the
raw-counts spectrum. The \LB\ line is fit with a model line profile
derived from raytraces of the telescope light path kindly provided by
J.~Kruk.  Because the \Osix\ lines are expected to fill the aperture,
the same profile is used for them as for \LB. The wavelength of the
\LB\ line is free to vary; those of the \Osix\ lines are tied to it
assuming that the emitting gas lies at the cluster redshift. The line
strengths also vary freely, except that the \Osix\ lines are held to
the 2:1 ratio expected from an optically thin plasma. The \Osix\ line
is then fixed at zero counts and the spectrum refit. If $\Delta \chi^2
> 4$ (corresponding to a \sig{2} deviation for one interesting
parameter, in this case the total counts in the \Osix\ line;
\citeNP{Avni76}), we claim a detection and determine \sig{1} error
bars by raising the model line strength above the best-fit level until
$\Delta \chi^2 = 1$.  Otherwise, we create a ``background-only'' data
set by subtracting the \Osix\ component of the best-fitting model
(which, though not statistically significant, is usually non-zero) from
the observed spectrum. We fit a linear continuum to the background
spectrum, then raise the model \Osix\ line strength above zero until
$\Delta \chi^2 = 4$ to set a \sig{2} limit on the flux of the
\Osix\ line. Limits to the \Cfour\ flux are set in the same way.
\fig{coma} shows the HUT spectrum of the Coma cluster in the region of
the redshifted \Osix\ feature; model spectra without \Osix\ and with
the line fixed at its \sig{2} upper limit are overplotted.

\begin{figure}
\plotone{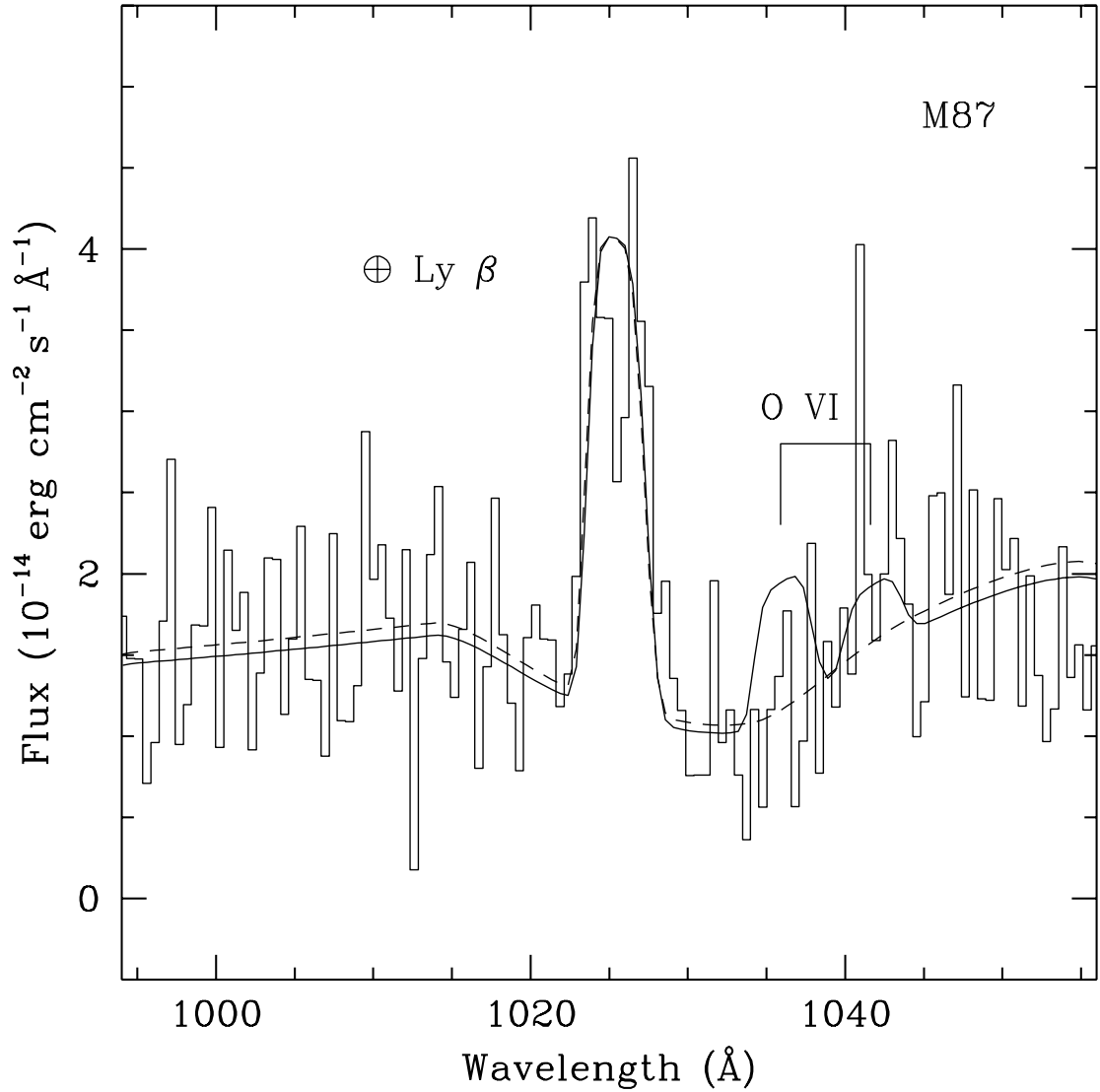}
\caption{HUT spectrum of M87 in the Virgo cluster, which
does not show significant \Osixwave\ emission. The data are shown as a
histogram; overplotted are models including no \Osix\ emission
(dotted line) and \Osix\ at our 2 $\sigma$ upper limit (solid line).
The fit to the stellar component is from Brown et al. (1995).
\label{m87}}
\end{figure}

A more complex model is required to reproduce the spectrum of M87,
which has a strong stellar component. As shown in \fig{m87}, the
redshifted \Osix\ line falls within the broad stellar \LB\ absorption
feature. In order to estimate the stellar contribution at this
wavelength, we flux-calibrate the data and model it with the synthetic
spectrum, based on the stellar atmosphere models of \citeN{Kurucz92}
and \citeN{CM87} and the evolutionary tracks of
\citeN{Dorman93} and \citeN{VW94}, originally fit
to these data by \citeN{BFD95}. The model spectrum is sampled at
10 \AA\ resolution; we interpolate it to 0.5 \AA\ using the spline
function in the IRAF routine INTERP. Limits on the flux of the \Osix\
doublet are then set as described above.

Our limits on the line fluxes are presented in Table \ref{limits}.
None of the spectra show significant FUV line emission; the relatively
high \Osix\ limit derived from the M87 spectrum reflects the greater
uncertainty in the spectral continuum. Dereddened intensities are
calculated assuming the values of \ebv\ in Table \ref{targets} and the
extinction parameterization of \citeANP{CCM89} (\citeyearNP{CCM89};
hereafter CCM). Though the FUV portion of the CCM extinction curve is
based on an extrapolation of data obtained at longer wavelengths, it is
able to reproduce the general shapes of the extinction curves of
early-type stars observed with {\it Copernicus} (CCM) and has been used
successfully to model the FUV spectra of hot stars and galaxies
observed with HUT (\eg, \shortciteNP{BFD95}; \citeNP{DDF95}). We
follow \shortciteN{BFD95} in assuming $R_V = 3.05$ for the diffuse
ISM along these lines of sight (\cf\ \citeNP{Buss94}); uncertainties
in $R_V$ do not greatly affect our result, as the FUV extinction
A$(\lambda)$ varies slowly with $R_V$ when \ebv\ is small.

\section{Discussion}

We do not detect significant emission from redshifted \Osixwave\ and
\Cfourwave\ in the HUT spectra of five clusters of galaxies. Are our
limits consistent with the EUV emission detected in the Virgo Cluster?

\citeANP{LMBLHS96} assume a distance to M87 of 20.6 Mpc.  Fitting a
two-temperature model to the combined \euve\ and \rosat\ data,
they derive the temperature, emission integral
($EI \equiv \int n_p n_e dV$), and abundance of both the warm and
hot components of the ICM as a function of radius. For the warm
component in the central region ($r < 3\arcmin$) of the cluster, they
find a temperature of $5.0^{+1.9}_{-} \times 10^5$~K, an emission
integral of $3.71^{+}_{-2.09} \times 10^{64}$ cm$^{-3}$, and an
abundance of $0.592^{+0.037}_{-0.061}$ relative to solar. (The lower
bound of the temperature and the upper bound of the emission integral
are unconstrained in the fit.) Using the best-fit values for each of
these parameters and line emissivities estimated with the
Raymond-Smith \citeyear{RS77} plasma code by C.-Y. Hwang, we find that the
\Osix\ flux expected from the warm gas in the central region of the Virgo
cluster is $6.9 \times 10^{-12}$ \ERGS. If we assume that the
\Osix\ flux is evenly distributed within this region, then its surface
brightness is $2.9 \times 10^{-6}$ \INTENS, or 60\% of our dereddened
upper limit.

The reddening to M87 may be less than we have assumed. The value
$E(B-V) = 0.023$ presented in Table \ref{targets} \cite{BH84} is
derived from the \ion{H}{1} map of \citeN{Heiles75}, which gives $N_H =
3.7 \times 10^{20}$ cm$^{-2}$ in the direction of M87.
\citeANP{LMBLHS96} have re-mapped the Galactic \ion{H}{1} column
density near M87 and find $N_H \sim (1.8-2.1) \times 10^{20}$
cm$^{-2}$, which corresponds to $E(B-V) = 0.000 \pm 0.015$ according to
the prescription of \citeANP{BH84} \citeyear{BH82,BH84}. \citeN{GHJN94}
combine deep CCD images in $B$, $V$, $I$ and narrow-band
\HA$+$[\ion{N}{2}] to derive the amount and morphology of ionized gas
and dust in 56 elliptical galaxies. They find no evidence for dust in
M87, save for that associated with the \HA$+$[\ion{N}{2}] jet near the
nucleus \cite{Jarvis90}, to a detection limit of $A_B \sim 0.02$, or
$E(B-V) \sim 0.005$. If extinction along the line of sight to M87 is
indeed negligible, then we may compare the predicted \Osix\ surface
brightness with the observed limit presented in Table \ref{limits},
rather than the dereddened value.

It is likely that the emission integral of the warm gas---and thus the
predicted \Osix\ surface brightness---is not constant across the
central 6\arcmin\ of the cluster. \citeN{SGF82} find that the
surface brightness of the hot component of the ICM, as measured with
the {\it Einstein} HRI, falls off as $r^{-1.1}$ for $20\arcsec < r <
4\arcmin$. The ratio of the EUV excess to the x-ray flux in the Virgo
cluster is roughly constant with radius (Lieu, private communication).
If we assume that the warm gas has a profile similar to that of the HRI
data (and for simplicity that it is flat for $r < 20\arcsec$), then the
fraction of the predicted \Osix\ flux falling in the HUT slit rises by
a factor of 5.6, to a mean surface brightness of $1.6 \times 10^{-5}$
\INTENS, more than four times the observed (and three times the
dereddened) \Osix\ limit set with \hut.

Our upper limit to the \Osix\ flux corresponds to a cooling rate within
the \hut\ aperture of $\dot{M} \lesssim 2.4 M_{\sun}$ yr$^{-1}$,
based on the models of \citeN{EC86} rescaled
by the abundance ($Z = 0.592$ $Z_{\sun}$) of the warm gas. This
value is consistent with the rate $\dot{M} \sim 2 M_{\sun}$ yr$^{-1}$
derived by \citeN{SCFN84} for the hot component of the ICM
within $\sim$ 1\arcmin\ of the galactic center.

\begin{figure}
\plotone{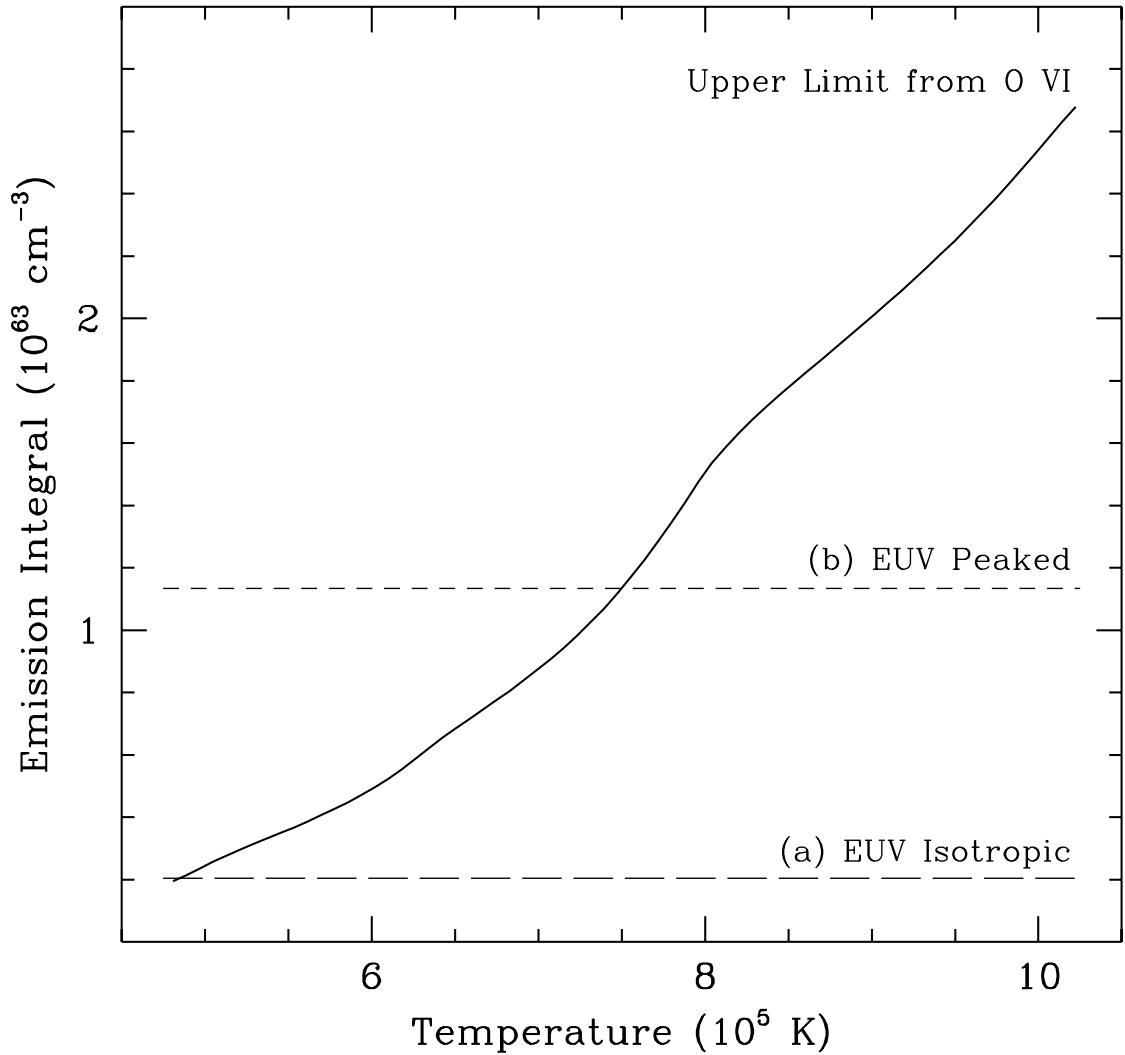}
\caption{Upper limit to the emission integral ($EI \equiv
\int n_p n_e dV$) as a function of gas temperature (solid line).  This
curve is derived from the observed (\ie, not dereddened) upper limit to
the \Osix\ flux in the HUT spectrum of M87. The horizontal lines
indicate the best-fit value of the emission integral derived from the
\euve\ data for the central region ($r < 3\arcmin$) of M87 scaled to the
area of the HUT slit assuming that the EUV surface brightness (a) is
constant across the inner region of the cluster (long dashed line),
and (b) scales as the x-ray surface brightness (short dashed line).
For case (b), the HUT data impose a lower limit to the gas temperature
of $7.5 \times 10^5$ K.
\label{ei_temp}}
\end{figure}

A substantial reservoir of warm gas may reside within a cluster of
galaxies if its temperature is greater than we have assumed.
The \Osix\ line emissivity depends critically on the gas temperature,
falling by a factor of 10 as the temperature rises from 5 to 10 $\times
10^5$~K. From the observed limit to the \Osix\ flux of M87, we
derive an upper limit to the emission integral as a function of
temperature. In \fig{ei_temp}, we compare this limit with the best-fit
value of the emission integral derived from the \euve\ data by
\citeANP{LMBLHS96}, scaled to the area of the HUT slit assuming that
the EUV surface brightness (a) is constant across the inner region of
the cluster, and (b) scales as the x-ray surface brightness. The value
of the emission integral corresponding to case (b), $EI = 1.1 \times
10^{63}$ cm$^{-3}$, is higher than the upper limit derived from the HUT
data unless $T \gtrsim 7.5 \times 10^5$~K.

\citeANP{LMBLHS96} report that the \rosat\ data for Coma and the Abell
clusters are best modeled by a two-temperature gas in which the cooler
component has $T \sim 10^6$~K. The lower \Osix\ emissivity expected of
a $10^6$~K gas may reconcile the presence of substantial warm gas with
the limits set by the HUT data. As more EUV and X-ray observations of
these clusters become available, the \Osix\ limits presented in Table
\ref{limits} will provide important constraints on models of their
intracluster media.

\section{Conclusions}

We have searched the spectra of five clusters of galaxies observed with
the Hopkins Ultraviolet Telescope for redshifted \Osixwave\ and
\Cfourwave\ emission. We do not detect significant emission in either
feature in any of the spectra.  If the surface brightness of the EUV
emission in the central region ($r < 3\arcmin$)  of the Virgo cluster
is centrally condensed, then our upper limit to the
\Osix\ surface brightness is inconsistent with that predicted for the
$5 \times 10^5$~K gas proposed by \citeANP{LMBLHS96} This
inconsistency may be alleviated if the gas temperature is $\gtrsim 7.5
\times 10^5$~K.  HUT limits on the \Osix\ surface brightness of the
four other clusters can provide important constraints on models of
their intracluster media.

\acknowledgments

We would like to thank J.~Kruk for providing raytrace results
and C.-Y.~Hwang for calculating line emissivities. This research has made
use of the NASA/IPAC Extragalactic Database (NED), which is operated by
the Jet Propulsion Laboratory, Caltech, under contract with the
National Aeronautics and Space Administration, the NASA ADS Abstract
Service, and the Catalogue Service of the CDS, Strasbourg, France. We
wish to acknowledge the efforts of our colleagues on the HUT team as
well as the many NASA personnel who helped make the Astro-2 mission
successful. The Hopkins Ultraviolet Telescope Project is supported by
NASA contract NAS 5-27000 to the Johns Hopkins University. W.V.D. and
M.H. acknowledge support from NASA grant NAG5-696.


\begin{deluxetable}{lccccccc}
\tablewidth{0pc}
\tablecaption{Target Summary\label{targets}}
\tablecolumns{8}
\tablehead{
\colhead{Name}   &
\colhead{\it l}          & \colhead{\it b}  &
\colhead{\it z}          & \colhead{\ebv\tablenotemark{a}}  &
\colhead{AMET\tablenotemark{b}}          &
\colhead{Time (s)}       & \colhead{Slit\tablenotemark{c}}
}
\startdata
Hercules & \phn31.6 & 44.5 & 0.0369\phn & 0.015 & 205.8 & 1592 & 6 \nl
& & & & & 277.4 & 1592 & 6\nl
A1795   & \phn33.8 & 77.2 & 0.0616\phn & 0.000 & 278.8 & 2292 & 6 \nl
Coma\tablenotemark{d}  & \phn58.1 & 88.0 & 0.0232\phn & 0.013 & 253.0 & 1748 & 7 \nl
& & & & & 254.5 & 2004 & 7 \nl
A1367   &    234.8 & 73.0 & 0.0216\phn & 0.000 & 179.8 & 1928 & 7 \nl
M87     &    283.8 & 74.5 & 0.00428 & 0.023 & 286.8 & \phn954 & 6 \nl
\enddata
\tablenotetext{a}{From \citeNP{BH84}.}
\tablenotetext{b}{Start time of observation in hours from launch (1995:061:06:38:13 UT).}
\tablenotetext{c}{Slit 6 measures 10\arcsec $\times$56\arcsec;
slit 7 is 20\arcsec\ in diameter.}
\tablenotetext{d}{The two Coma pointings were offset by approximately 2\farcm4.}
\end{deluxetable}

\newpage

\begin{deluxetable}{lcccc}
\tablewidth{0pc}
\tablecaption{2 $\sigma$ Upper Limits to Far-UV Emission Lines \label{limits}}
\tablecolumns{5}
\tablehead{
\multicolumn{1}{c}{} &
\multicolumn{2}{c}{\ion{O}{6} $\lambda \lambda 1032,1038$} &
\multicolumn{2}{c}{\ion{C}{4} $\lambda \lambda 1548,1551$} \\
\colhead{Name}    &
\colhead{Observed}	& \colhead{Dereddened}	&
\colhead{Observed}	& \colhead{Dereddened}
}
\startdata
Hercules & 0.577 & 0.700 & 0.565 & 0.631 \nl
A1795   & 1.12\phn & 1.12\phn & 1.19\phn & 1.19\phn \nl
Coma  & 1.21\phn & 1.43\phn & 1.65\phn & 1.81\phn \nl
A1367   & 2.24\phn & 2.24\phn & 1.20\phn & 1.20\phn \nl
M87      & 3.52\phn & 4.84\phn & 3.07\phn & 3.64\phn \nl
\enddata
\tablecomments{Units are $10^{-6}$ erg cm$^{-2}$ s$^{-1}$ sr$^{-1}$.
Dereddened intensities assume a CCM extinction curve, the
\ebv\ value from Table \ref{targets}, and $R_V = 3.05$.
}
\end{deluxetable}

\clearpage


\bibliographystyle{apj}
\bibliography{apjmnemonic,bib}

\begin{thebibliography}{}

\bibitem[\protect\citeauthoryear{{Avni}}{{Avni}}{1976}]{Avni76}
{Avni}, Y. 1976, ApJ, 210, 642

\bibitem[\protect\citeauthoryear{{B\"ohringer} et~al.}{{B\"ohringer}
  et~al.}{1995}]{Boh95}
{B\"ohringer}, H., {Nulsen}, P. E.~J., {Braun}, R.,  \& {Fabian}, A.~C. 1995,
  MNRAS, 274, L67

\bibitem[\protect\citeauthoryear{{Brown}, {Ferguson}, \& {Davidsen}}{{Brown}
  et~al.}{1995}]{BFD95}
{Brown}, T.~M., {Ferguson}, H.~C.,  \& {Davidsen}, A.~F. 1995, ApJ, 454, L15

\bibitem[\protect\citeauthoryear{{Burstein} \& {Heiles}}{{Burstein} \&
  {Heiles}}{1982}]{BH82}
{Burstein}, D.,  \& {Heiles}, C. 1982, AJ, 87, 1165

\bibitem[\protect\citeauthoryear{{Burstein} \& {Heiles}}{{Burstein} \&
  {Heiles}}{1984}]{BH84}
{Burstein}, D.,  \& {Heiles}, C. 1984, ApJS, 54, 33

\bibitem[\protect\citeauthoryear{{Buss} et~al.}{{Buss} et~al.}{1994}]{Buss94}
{Buss}, R.~H.,~Jr., {Allen}, M., {McCandlis}, S., {Kruk}, J., {Liu}, J.-C.,  \&
  {Brown}, T. 1994, ApJ, 430, 630

\bibitem[\protect\citeauthoryear{{Cardelli}, {Clayton}, \& {Mathis}}{{Cardelli}
  et~al.}{1989}]{CCM89}
{Cardelli}, J.~A., {Clayton}, G.~C.,  \& {Mathis}, J.~S. 1989, ApJ, 345,
  245~(CCM)

\bibitem[\protect\citeauthoryear{{Clegg} \& {Middlemass}}{{Clegg} \&
  {Middlemass}}{1987}]{CM87}
{Clegg}, R. E.~S.,  \& {Middlemass}, D. 1987, MNRAS, 228, 759

\bibitem[\protect\citeauthoryear{{Davidsen} et~al.}{{Davidsen}
  et~al.}{1992}]{HUT_INSTR}
{Davidsen}, A.~F., et~al. 1992, ApJ, 392, 264

\bibitem[\protect\citeauthoryear{{Dixon}, {Davidsen}, \& {Ferguson}}{{Dixon}
  et~al.}{1995}]{DDF95}
{Dixon}, W.~V., {Davidsen}, A.~F.,  \& {Ferguson}, H.~C. 1995, ApJ, 454, L47

\bibitem[\protect\citeauthoryear{{Dorman}, {Rood}, \& {O'Connell}}{{Dorman}
  et~al.}{1993}]{Dorman93}
{Dorman}, B., {Rood}, R.~T.,  \& {O'Connell}, R.~W. 1993, ApJ, 419, 596

\bibitem[\protect\citeauthoryear{{Edgar} \& {Chevalier}}{{Edgar} \&
  {Chevalier}}{1986}]{EC86}
{Edgar}, R.~J.,  \& {Chevalier}, R.~A. 1986, ApJ, 310, L27

\bibitem[\protect\citeauthoryear{{Fabian}}{{Fabian}}{1996}]{Fabian96}
{Fabian}, A.~C. 1996, Science, 271, 1244

\bibitem[\protect\citeauthoryear{{Forman} \& {Jones}}{{Forman} \&
  {Jones}}{1982}]{FJ82}
{Forman}, W.,  \& {Jones}, C. 1982, ARA\&A, 20, 547

\bibitem[\protect\citeauthoryear{{Goudfrooij} et~al.}{{Goudfrooij}
  et~al.}{1994}]{GHJN94}
{Goudfrooij}, P., {Hansen}, L., {J{\o}rgensen}, H.~E.,  \&
  {N{\o}rgaard-Nielsen}, H.~U. 1994, A\&AS, 105, 341

\bibitem[\protect\citeauthoryear{{Heiles}}{{Heiles}}{1975}]{Heiles75}
{Heiles}, C. 1975, A\&AS, 20, 37

\bibitem[\protect\citeauthoryear{{Jarvis}}{{Jarvis}}{1990}]{Jarvis90}
{Jarvis}, B.~J. 1990, A\&A, 240, L8

\bibitem[\protect\citeauthoryear{{Kriss}}{{Kriss}}{1994}]{Kriss94}
{Kriss}, G.~A. 1994, in ASP Conf. Ser. 61, Astronomical Data Analysis Software
  and Systems III, ed. D.~R. {Crabtree}, R.~J. {Hanisch}, \& J.~{Barnes} (San
  Francisco: ASP), 437

\bibitem[\protect\citeauthoryear{{Kruk} et~al.}{{Kruk}
  et~al.}{1995}]{HUT_INSTR2}
{Kruk}, J.~W., {Durrance}, S.~T., {Kriss}, G.~A., {Davidsen}, A.~F., {Blair},
  W.~P., {Espey}, B.~R.,  \& {Finley}, D. 1995, ApJ, 454, L1

\bibitem[\protect\citeauthoryear{{Kurucz}}{{Kurucz}}{1992}]{Kurucz92}
{Kurucz}, R.~L. 1992, in IAU Symposium No. 149, The Stellar Populations of
  Galaxies, ed. B.~{Barbuy} \& A.~{Renzini} (Dordrecht: Kluwer), 225

\bibitem[\protect\citeauthoryear{{Lieu} et~al.}{{Lieu} et~al.}{1996}]{LMBLHS96}
{Lieu}, R., {Mittaz}, J. P.~D., {Bowyer}, S., {Lockman}, F.~J., {Hwang}, C.-Y.,
   \& {Schmitt}, J. H. M.~M. 1996, ApJ, 458, L5

\bibitem[\protect\citeauthoryear{{Raymond} \& {Smith}}{{Raymond} \&
  {Smith}}{1977}]{RS77}
{Raymond}, J.~C.,  \& {Smith}, B.~W. 1977, ApJS, 35, 419

\bibitem[\protect\citeauthoryear{{Schreier}, {Gorenstein}, \&
  {Feigelson}}{{Schreier} et~al.}{1982}]{SGF82}
{Schreier}, E.~J., {Gorenstein}, P.,  \& {Feigelson}, E.~D. 1982, ApJ, 261, 42

\bibitem[\protect\citeauthoryear{{Stewart} et~al.}{{Stewart}
  et~al.}{1984}]{SCFN84}
{Stewart}, G.~C., {Canizares}, C.~R., {Fabian}, A.~C.,  \& {Nulsen}, P. E.~J.
  1984, ApJ, 278, 536

\bibitem[\protect\citeauthoryear{{Vassiliadis} \& {Wood}}{{Vassiliadis} \&
  {Wood}}{1994}]{VW94}
{Vassiliadis}, E.,  \& {Wood}, P.~R. 1994, ApJS, 92, 125

\bibitem[\protect\citeauthoryear{{Voit}, {Donahue}, \& {Slavin}}{{Voit}
  et~al.}{1994}]{VDS94}
{Voit}, G.~M., {Donahue}, M.,  \& {Slavin}, J.~D. 1994, ApJS, 95, 87

\end{thebibliography}


\newpage


\end{document}